\documentclass[aps,pre,reprint,noshowpacs,longbibliography,superscriptaddress,twocolumn,notitlepage]{revtex4-1}
\usepackage{graphicx}
\usepackage{amsmath,amsfonts}
\usepackage[pdftex]{hyperref}
\usepackage{bm}
\usepackage[caption=false]{subfig}
\begin{document}
%
%
%
%
%
%
\title{Percolation and the effective structure of complex networks}
\author{Antoine \surname{Allard}}
\affiliation{Departament de F\'isica de la Mat\`eria Condensada and Institute of Complex Systems (UBICS), Universitat de Barcelona, Mart\'i i Franqu\`es 1, E-08028 Barcelona, Spain}
\affiliation{D\'epartement  de  physique,  de  g\'enie  physique et  d'optique, Universit\'e  Laval,  Qu\'ebec  (Qu\'ebec),  Canada  G1V  0A6}
\author{Laurent \surname{H\'ebert-Dufresne}}
\affiliation{Department of Computer Science and Vermont Complex Systems Center, University of Vermont, Burlington VT, USA}
\affiliation{D\'epartement  de  physique,  de  g\'enie  physique et  d'optique, Universit\'e  Laval,  Qu\'ebec  (Qu\'ebec),  Canada  G1V  0A6}
\date{\today}
\begin{abstract}
  Analytical approaches to model the structure of complex networks can be distinguished into two groups according to whether they consider an intensive (e.g., fixed degree sequence and random otherwise) or an extensive (e.g., adjacency matrix) description of the network structure.  While extensive approaches---such as the state-of-the-art Message Passing Approach---typically yield more accurate predictions, intensive approaches provide crucial insights on the role played by any given structural property in the outcome of dynamical processes.  Here we introduce an intensive description that yields almost identical predictions to the ones obtained with MPA for bond percolation.  Our approach distinguishes nodes according to two simple statistics: their degree and their position in the core-periphery organization of the network.  Our near-exact predictions highlight how accurately capturing the long-range correlations in network structures allows to easily and effectively compress real complex network data.
\end{abstract}
\maketitle
%
%
%
%
%
%
The structure of real complex networks lies somewhere in-between order and randomness~\cite{Watts1998,Goldenfeld1999,Strogatz2005}, with the consequence that it cannot typically be fully characterized by a concise set of synthesizing observables.  This \textit{irreductibility} explains why most theoretical approaches to model complex networks are inspired by statistical physics in that they consider ensembles of networks constrained by the values of observables (e.g. density of links, degree-degree correlations, clustering coefficient, degree/motif distribution) and otherwise organized randomly.  These approaches have three notable advantages. First, they usually yield analytical treatment. Second, they are \textit{intensive} in network size, meaning that their complexity scales with the support of the observables (i.e., sub-linearly with the numbers of nodes and links). Third, they provide null models, of which many have led to the identification of fundamental properties characterizing the structure of real complex networks~\cite{Newman2010,Barabasi2016}.

Despite important leaps forward in recent years, these approaches still fail to capture enough information to systematically provide accurate quantitative predictions of most dynamical processes on real complex networks.  The reason for this shortcoming is that the properties from which the ensembles are constructed are not constraining enough; the ensembles are ``too large'' such that the original real networks are exceptions, rather than typical instances, in the ensembles.  As a result, the current state-of-the-art approach---the so-called \textit{message passing approach} (MPA)~\cite{Karrer2014}---requires the whole structure to be specified as an input (i.e., the adjacency matrix, or a transformation thereof).  This method is interesting because it is mathematically principled, meaning that it yields \textit{exact} results on trees, and offers inexact, albeit generally good, predictions on networks containing loops (i.e., most real complex networks)~\cite{Radicchi2015}.

However, by considering the whole structure of networks and thereby considering every link on equal footing, the accuracy of the MPA comes at a significant computational and conceptual cost.  First, its time and space complexity are \textit{extensive} in the number of links and therefore in the size of the network.  Second, and most importantly, it does not provide any insight on the role played by any given structural property in the outcome of a dynamical process.  With the MPA, getting good predictions comes at the expense of understanding what led to that outcome.

In this paper, we bridge the gap between intensive and extensive approaches to the mathematical modeling of bond percolation on networks.  We introduce a random network ensemble that relies solely on an \textit{intensive} description of the network structure that, nevertheless, yields predictions that are comparable to the ones from the MPA for most of the 111 real complex networks considered in this study.  This ensemble is based on the \textit{onion decomposition} (OD), a refined $k$-core decomposition~\cite{Hebert-Dufresne2016a}.  Critically, the OD can be translated into local connection rules allowing an exact mathematical treatment using probability generating functions (pgf) in the limit of large network size.  This approach leads to exact predictions on trees like the MPA, and highlights the critical contribution of the OD to an accurate effective mathematical description of real complex networks.
\begin{figure}[t]
  \begin{center}
    \includegraphics[width=\linewidth]{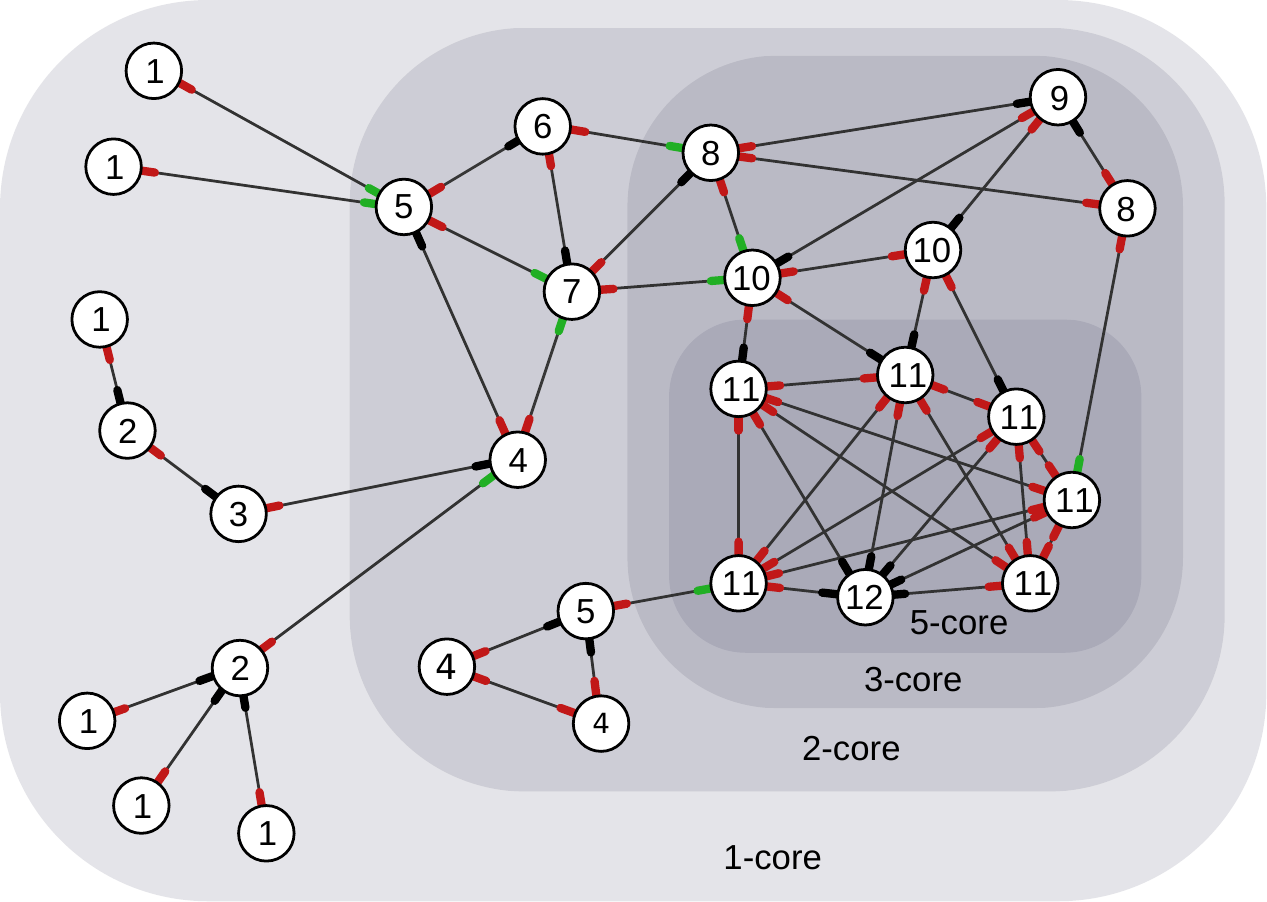}
    \caption{Illustration of the Onion Decomposition (OD) of a simple network.  The number of the layer to which each node belongs is indicated and the different $k$-cores are shown using increasingly darker background shades.  The color of each stub according the LCCM is also shown.}
    \label{fig:illutrationOD}
  \end{center}
\end{figure}
%
%
%
%
%
\section*{Results and discussions}
%
Most analytical models of complex networks rely on some variation of the tree-like approximation which assumes that complex networks have essentially no loops beyond some local structure of interest~\cite{Karrer2010,Allard2015}.  While this approximation is inaccurate for the vast majority of real complex networks, it nevertheless allows an elegant mathematical treatment which typically works surprisingly well~\cite{Melnik2011}.  In the case of the MPA, the tree-like approximation implies that a lot of information given to the model is thrown away due to loops being included in the input information (i.e., the adjacency matrix) to then be mathematically ignored.  We here propose to limit the information we give to our model by compressing complex networks following their tree-like decomposition.  We therefore rely on a known peeling process, which iteratively removes leaves (i.e., the peripheral nodes of the network) to calculate the depth of every node in the \textit{effective tree}.

Taking this information into account, we then focus on predicting the outcome of bond percolation on complex networks: a canonical problem of network science analogous to many applied problems such as disease propagation or network resilience~\cite{Latora2017}.  Given a network structure, this simple stochastic process consists in the occupation of each original link with probability $p$.  We aim to predict the size of the largest connected component composed of occupied links, $S$, as well as the percolation threshold, $p_\mathrm{c}$, above which that component corresponds to a macroscopic fraction of the network.  The outcome of percolation depends on structural properties at all scales, thus making it a good benchmark for theoretical network models.
\begin{figure*}[t]
  \begin{center}
    \subfloat[Message Passing Approach]{\centering \includegraphics[width=.27\textwidth, angle=0]{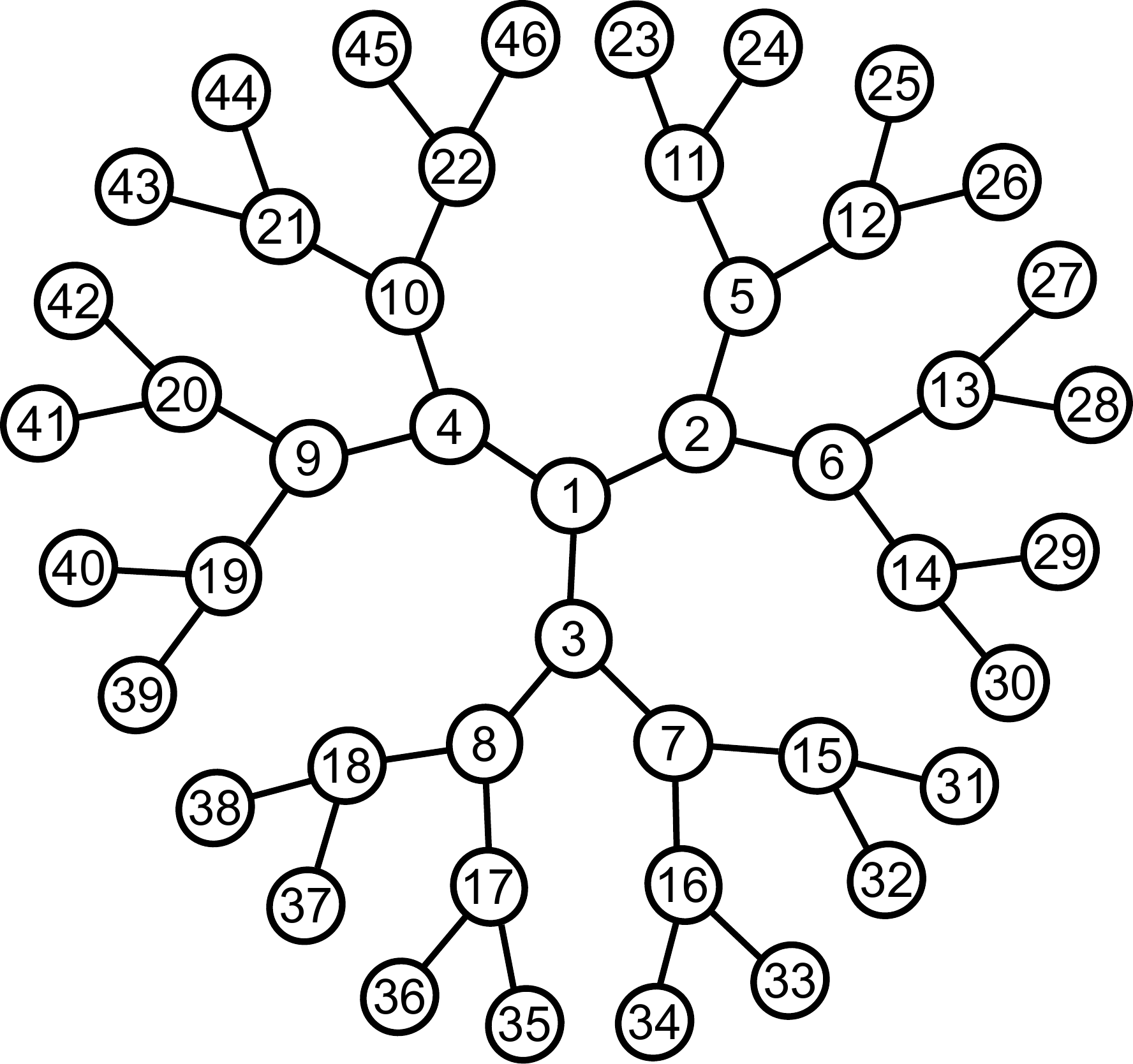}}\hspace{0.75cm}
    \subfloat[Layered and Correlated Configuration Model]{\centering \includegraphics[width=.27\textwidth, angle=0]{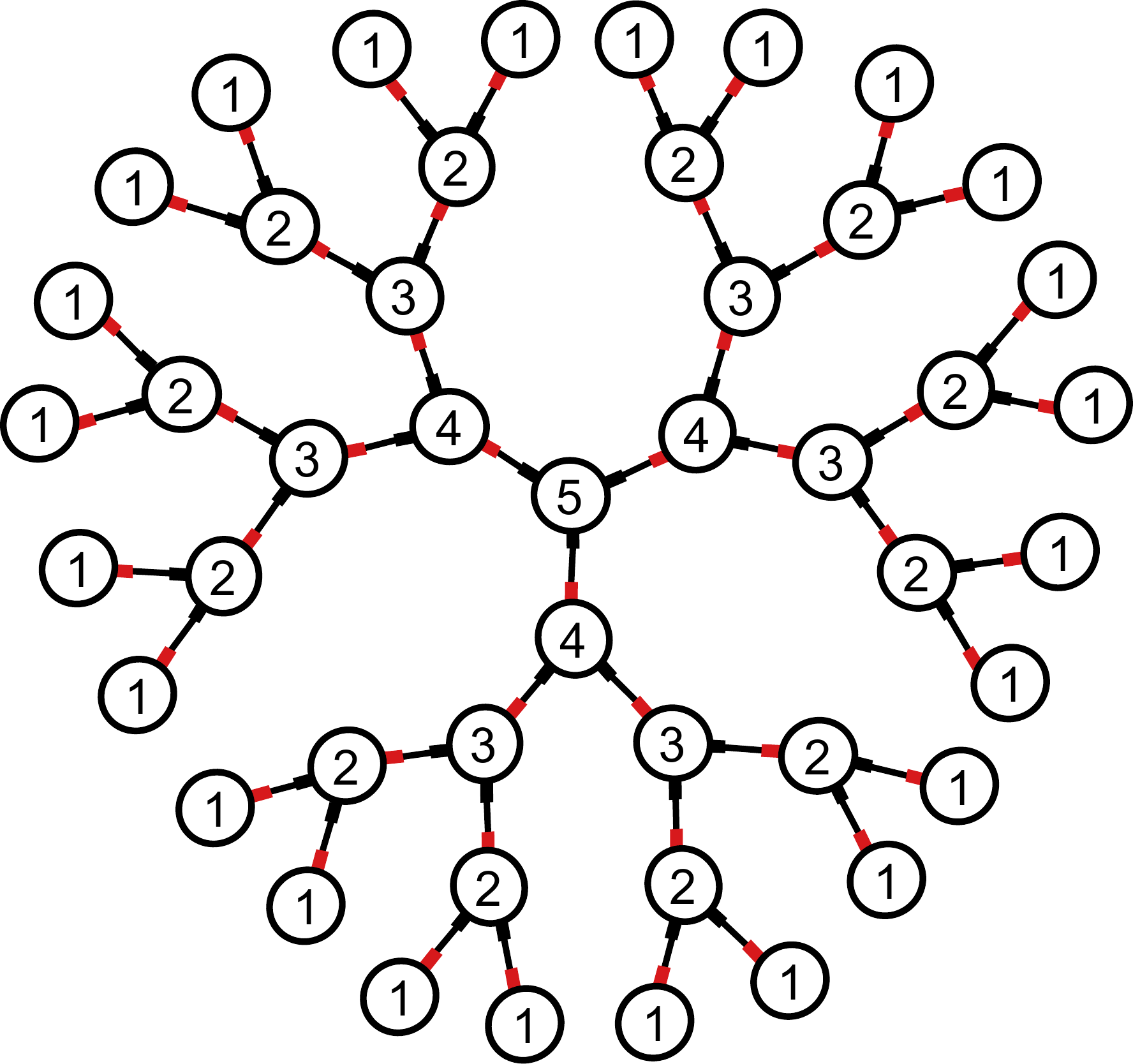}}\hspace{0.75cm}
    \subfloat[Configuration Model]{\centering \includegraphics[width=.27\textwidth, angle=0]{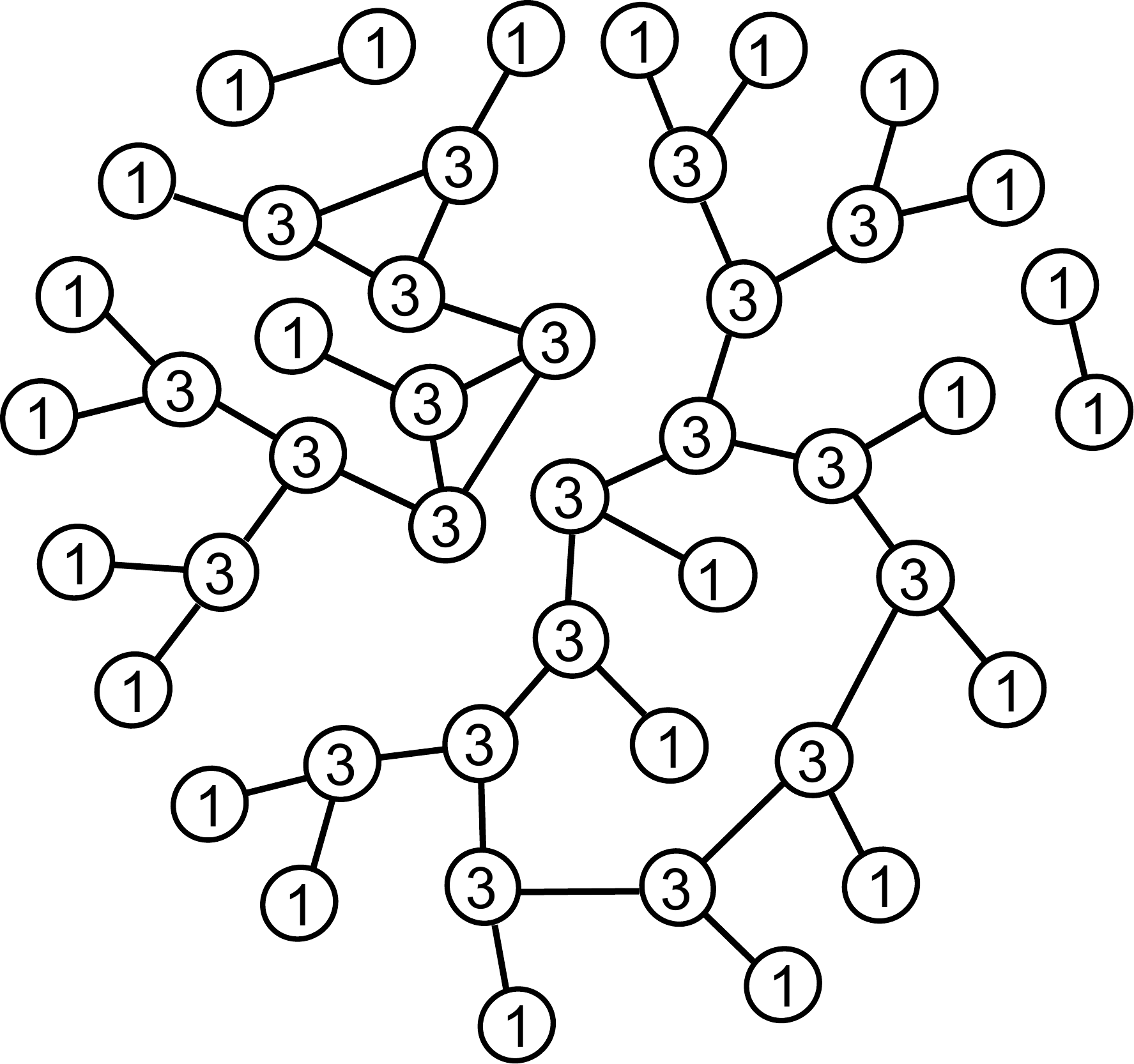}}
    \caption{Compression of a perfect tree with different network models. (a) The Message Passing Approach assigns a unique ID to every node and preserves the full structure of the tree. (b) The Layered and Correlated Configuration Model assigns an ID to every node corresponding to its degree and its position in the core-periphery structure of the network.  Degrees are not shown to lighten the presentation.  Stubs are colored according to the layer to which they point: red if they point to more central layers and black if they point to the previous layer.  There are no green stubs in this example.  (c) The Configuration Model assigns an ID to every node according to its degree before randomly connecting them, therebyy destroying the mesoscopic and macroscopic structure of the original network. The Correlated Configuration Model fixes the number of links between different degree classes, and would therefore prohibit components formed by two nodes with degree 1, but would otherwise be very similar to the configuration model shown here.}
    \label{fig:trees}
  \end{center} 
\end{figure*}
%
%
%
%
%
\subsection*{Onion decomposition}
%
\begin{figure*}[t] 
  \centering
  \includegraphics[width = \textwidth]{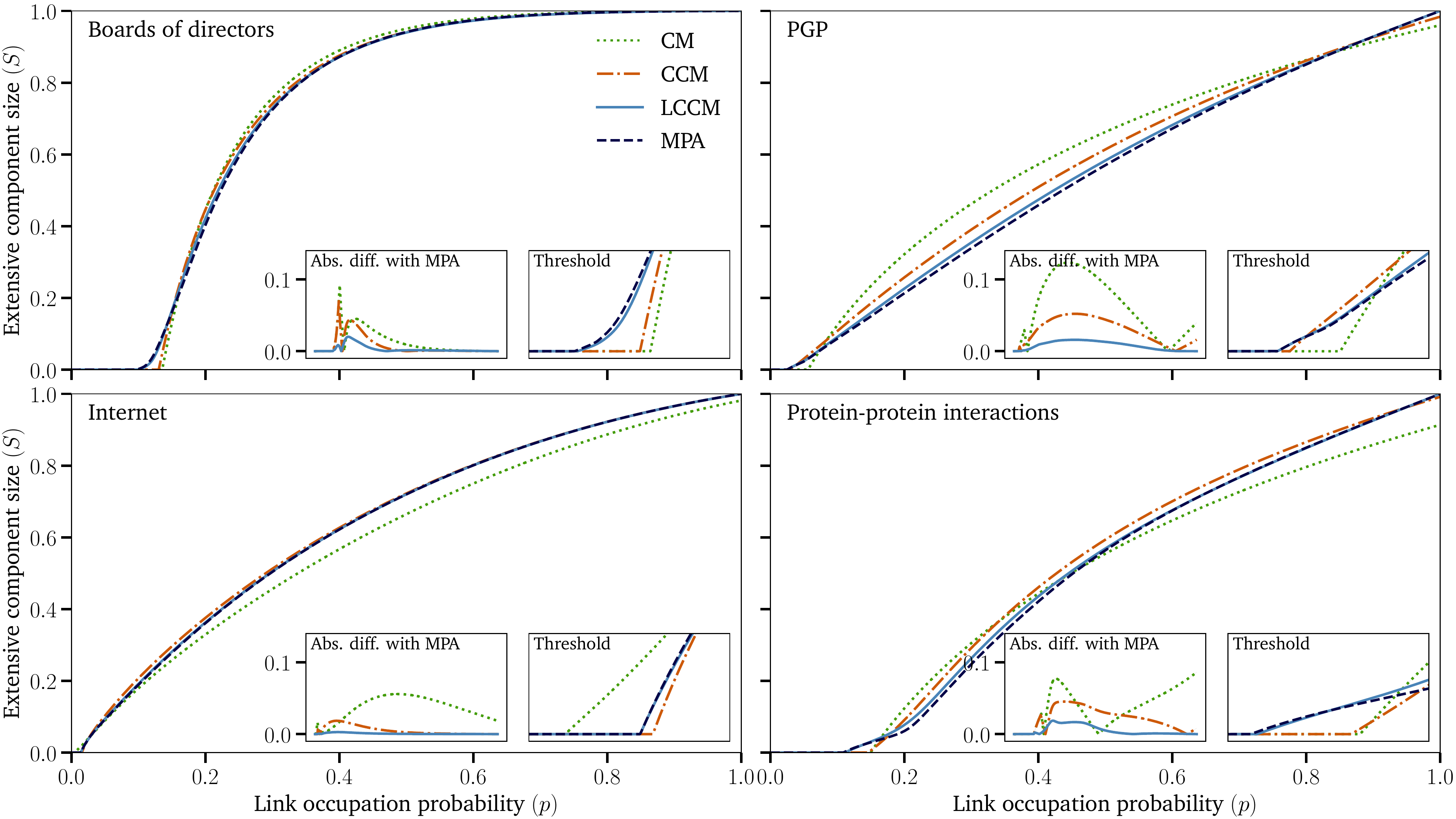}
  \caption{Relative size of the extensive components predicted by the LCCM with the CM, the CCM and the MPA for 4 representative real network datasets. (upper left) One-mode projection of a Norwegian boards of directors bipartite network~\cite{Seierstad2011}. (upper right) PGP web of trust~\cite{Boguna2004}. (lower left) A subset of the Internet at the autonomous level~\cite{Leskovec2005}. (lower right) Protein-protein interaction network of Homo sapiens~\cite{Song2005}.  The insets show the absolute value of the difference between the MPA and the CM the CCM and the LCCM as a function of $p$, as well as an enlargement of the region around the percolation threshold.  The largest connected component was used for all dataset.}
  \label{fig:bifurcation}
\end{figure*}
The $k$-core decomposition is a well-known network metric that identifies a set of nested maximal sub-networks---the $k$-cores---in which each node shares at least $k$ links with the other nodes~\cite{Seidman1983,Dorogovtsev2006}.  A node belonging to the $k$-core but not to the $(k+1)$-core is said to be of \textit{coreness} $k$ and to be part of the $k$-\textit{shell}.  Nodes with a high coreness are generally seen as more central whereas nodes with low corenesses are seen as being part of the periphery of the network.  The onion decomposition (OD) refines the $k$-core decomposition by assigning a layer $l$ to each node to further indicate its \textit{position} within its shell (e.g., in the middle of the layer or at its boundary). The OD therefore unveils the internal organization of each centrality shell and, unlike the original $k$-core decomposition, can be used to assess whether the structure of a core is more similar to a tree or to a lattice, among other things~\cite{Hebert-Dufresne2016a}.

The OD of a given network structure is obtained via the following pruning process (see Fig.~\ref{fig:illutrationOD}).  First we remove every nodes with the smallest degree, $k_\mathrm{min}$; the coreness of these nodes is equal to $k_\mathrm{min}$ and they are part of the first layer ($l=1$).  Removing these nodes may yield nodes whose \textit{remaining} degree is now equal to or smaller than $k_\mathrm{min}$; these nodes must also be removed, have a coreness of $k_\mathrm{min}$ as well, but are part of the second layer ($l=2$).  If removing nodes of the second layer yields new nodes with a remaining degree equal to or lower than $k_\mathrm{min}$, they will be part of the third layer ($l=3$), will have a coreness of $k_\mathrm{min}$ and will also be removed.  This process is repeated until no new nodes with a remaining degree equal to or lower than $k_\mathrm{min}$ are left.  We then update the value of $k_\mathrm{min}$ to reflect the lowest remaining degree and repeat this whole process until every node has been assigned a coreness and a layer (the layer number keeps increasing such that each layer corresponds to a unique coreness).

An efficient implementation of this procedure has a run-time complexity of $\mathcal{O}(L \log N)$, where $L$ and $N$ are respectively the number of links and nodes, which implies that the OD can be quickly obtained for virtually any real complex network~\cite{Hebert-Dufresne2016a}.  Most importantly, nodes belonging to a same layer are \textit{topologically similar} with regard to the mesoscale centrality organization of the network.  Because the layer of a node is only weakly related to its degree (i.e., the coreness of a node provides a lower bound to its degree), the pair layer-degree can therefore be used to indicate how well a node is connected, but also to indicate its ``topological position'' in the network. It therefore allows us to discriminate central nodes from peripheral ones which, based on their degree alone, would have otherwise been deemed identical.
%
%
%
%
%
\subsection*{Effective random network ensemble: the LCCM}
%
From the pruning process described above, it can be concluded that a node of coreness $c$ belonging to the $l$-th layer is in one of two scenarios. 1) It must have \textit{exactly} $c$ links to nodes in layers $l^\prime \geq l$ if layer $l$ is the first layer of the $c$-shell (i.e., nodes in layer $l-1$ belong to the $c^\prime$-shell with $c^\prime < c$). 2) Otherwise, if it is not in the first layer of its $c$-shell, it must have \textit{at least} $c+1$ links to nodes of layers $l^\prime \geq l-1$ and \textit{at most} $c$ links to nodes of layers $l^\prime \geq l$.  The distinction between the two scenarios is that nodes not in the first layer of their shell require at least one link to the previous layer to \textit{anchor} them to their own layer.  Also, the common feature of these scenarios is that a node of coreness $c$ needs at least $c$ links with nodes of equal or greater coreness.

By rewiring the links of a given network using a degree-preserving procedure~\cite{Coolen2009,Fosdick2016} while ensuring that the aforementioned rules are respected at all time, it is possible to explore the ensemble of all possible single networks with the same fixed layer-degree sequence (i.e., the sequence of every pairs $(l, k)$ in the original network).  Exactly preserving the layers---and thus the coreness of every nodes---is of critical significance since previous rewiring approaches could only approximately preserve the $k$-core decomposition \cite{Hebert-Dufresne2013}.

Additionally, the pair layer-degree assigned to each node can be used to enforce two-point correlations (i.e., the (layer-degree)--(layer-degree) correlations), thus reducing the size of a random network ensemble.  This correlated ensemble can be explored via a double link swap Markov chain method preserving both the layer-degree sequence and the number of links within and between every node classes (i.e., nodes with the same layer-degree).  One way to implement this method is by first choosing one link at random (e.g., joining nodes A and B) and then choosing another link at random (e.g., joining nodes C and D) among the links that are attached to at least one node whose layer-degree pair is the same at one of the two nodes connected by the first link (e.g., A and C have the same layer-degree)~\cite{Colomer-de-Simon2013}.  The two links are then swapped (e.g., A becomes connected to D and B to C) if no self-link or multi-link would be created.  Doing so ensures that that both the degree sequence and the two-point correlations are preserved at all time.  We call \textit{layered and correlated configuration model} (LCCM) the ensemble of maximally random networks with a given joint layer-degree sequence and (layer-degree)--(layer-degree) correlations.

Since it preserves both the degree sequence and the degree-degree correlations, the LCCM is a subset of two commonly used random network ensembles defined by the \textit{configuration model} (CM)~\cite{Newman2002} and the \textit{correlated configuration model} (CMM)~\cite{Vazquez2003}; the latter being known for its fair accuracy in many applications~\cite{Melnik2011}.  The LCCM, however, distinguishes itself from these models (and other variants) by enforcing a mesoscopic organization via the layers of the OD.  This feature has the critical advantage of making the LCCM a mathematically principled approached in the sense that it exactly preserves the structure of a wide variety of trees (see Fig.~\ref{fig:trees}).  As we show below, this mesoscopic information accounts for a significant portion of the missing gap between the predictions of the intensive configuration models and the extensive, current state-of-the-art MPA.
%
%
%
%
%
\subsection*{Percolation on the LCCM}
%
We adapt the approach of Ref.~\cite{Allard2015} to solve bond/site percolation on the LCCM in the limit of large network size.  This approach requires to specify 1) the classes of nodes, which here correspond to the distinct pairs layer-degree noted $(l,k)$, and 2) the colors of stubs (i.e., half-links), which in the LCCM are identified based on the layer $l^\prime$ of the neighboring node.  More precisely, from the connection rules stated in the previous section, the LCCM requires to keep track of the number of links that each node in each layer $l$ shares with nodes i) in layers $l^\prime \geq l$, ii) in layer $l^\prime = l - 1$ and iii) in layers $l^\prime < l - 1$.  We identify the corresponding half-links as red, black and green stubs, respectively.  For instance, a link between nodes in layers 3 and 5 consists in a red stub stemming out of the node in layer 3 paired with a green stub belonging to the node in layer 5.  Note that a link between two given layers can only consist in a unique pair of stub colors, and the only allowed combinations are red-red, red-black and red-green.

From the link correlation matrix $\mathbf{L}$, whose entries specify the fraction of links within and between every classes of nodes, we can derive the function (see Methods)
\begin{align} \label{eq:varphi_definition}
  \varphi_{lk}(\bm{x}) = \sum_{k^\mathrm{r} k^\mathrm{b} k^\mathrm{g}} P_{lk}(k^\mathrm{r},k^\mathrm{b},k^\mathrm{g})
                        [x_{lk}^\mathrm{r}]^{k^\mathrm{r}} [x_{lk}^\mathrm{b}]^{k^\mathrm{b}} [x_{lk}^\mathrm{g}]^{k^\mathrm{g}}
\end{align}
generating the probability $P_{lk}(k^\mathrm{r},k^\mathrm{b},k^\mathrm{g})$ that a node in class $(l,k)$ has $k^\mathrm{r}$ red stubs, $k^\mathrm{b}$ black stubs and $k^\mathrm{g}$ green stubs, given the connection rules of the LCCM.  From the same link correlation matrix, we can also derive the functions (see Methods)
\begin{align} \label{eq:gamma}
  \gamma_{lk}^\mathrm{\alpha}(\bm{x}) & = \sum_{l^\prime k^\prime} \sum_{\alpha^\prime\in\{\mathrm{r},\mathrm{b},\mathrm{g}\}} Q_{lk}^\alpha(l^\prime, k^\prime, \alpha^\prime) x_{l^\prime k^\prime}^{\alpha^\prime} \ ,
\end{align}
for every $\alpha\in\{\mathrm{r},\mathrm{b},\mathrm{g}\}$, generating the probability $Q_{lk}^\alpha(l^\prime, k^\prime, \alpha^\prime)$ that a stub of color $\alpha$ stemming of a node of class $(l,k)$ is attached to a stub of color $\alpha^\prime$ belonging to a node in class $(l^\prime,k^\prime)$.  Combining these two functions yields the pgf generating the distribution of the number of nodes of each class that are neighbors of a randomly chosen node of class $(l,k)$
\begin{align} \label{eq:g}
  g_{lk}(\bm{x}) = \varphi_{lk}(\bm{\gamma(\bm{x})}) \ .
\end{align}
Note that this pgf also includes the colors of the stub through which these neighors are connected to the node of class $(l,k)$.  Similarly, the number of such nodes that can be reached from a node of class $(l,k)$ that has itself been reached by one of its stubs of color $\alpha$ is
\begin{align} \label{eq:f}
  f_{lk}^\alpha(\bm{x}) = \frac{1}{\langle k^\alpha \rangle_{lk}} \left. \frac{\partial \varphi_{lk}(\bm{x^\prime})}{\partial x_{lk}^{\prime\alpha}} \right|_{\bm{x^\prime}=\bm{\gamma(\bm{x})}} \ ,
\end{align}
where $\langle k^\alpha \rangle_{lk} = \frac{\partial \varphi_{lk}(\bm{1})}{\partial x_{lk}^\alpha}$ is the average number of stubs of color $\alpha$ nodes of class $(l,k)$ have.
\begin{figure*}[t] 
  \centering
  \includegraphics[width = 0.48\linewidth]{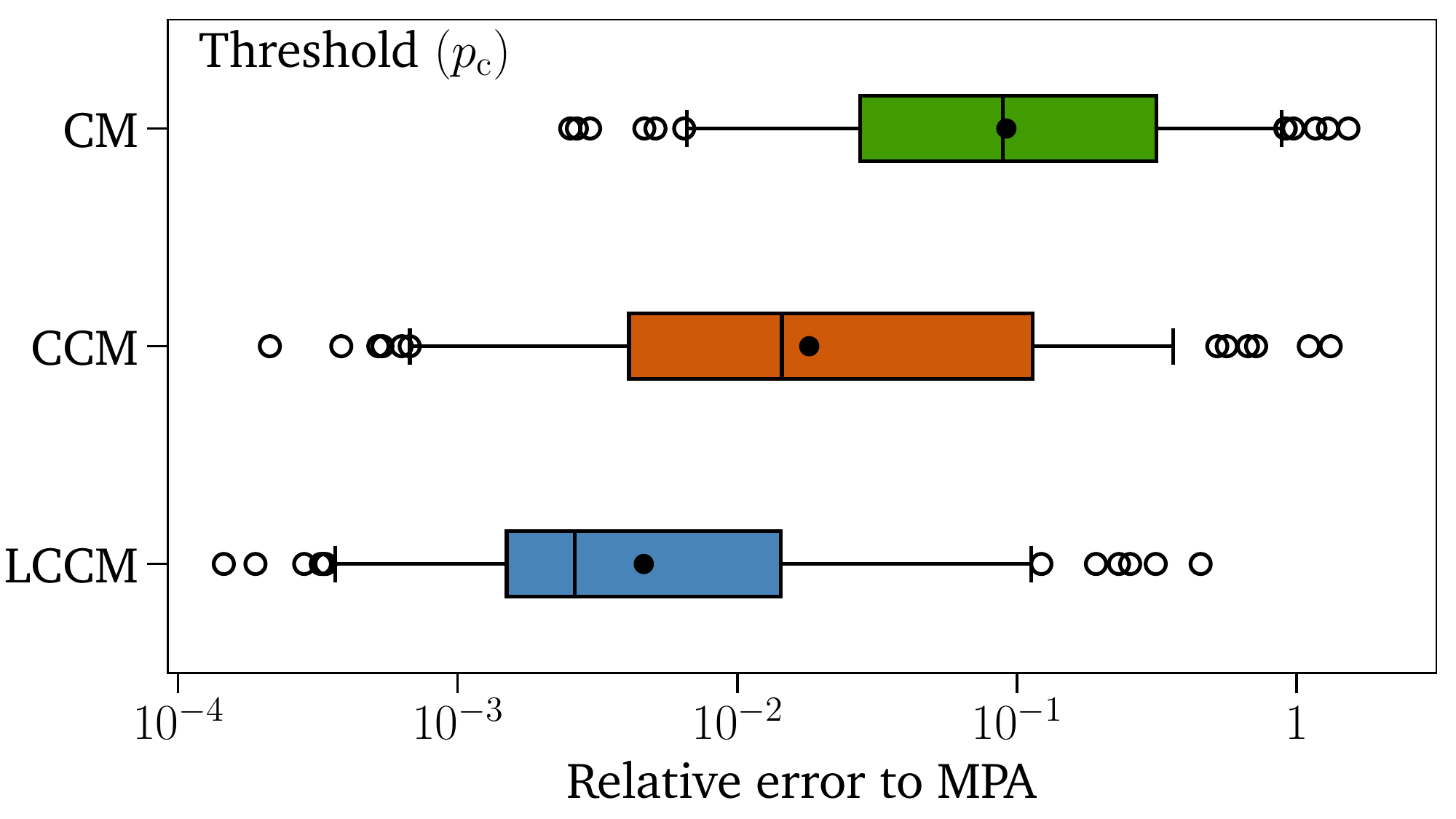}
  \includegraphics[width = 0.48\linewidth]{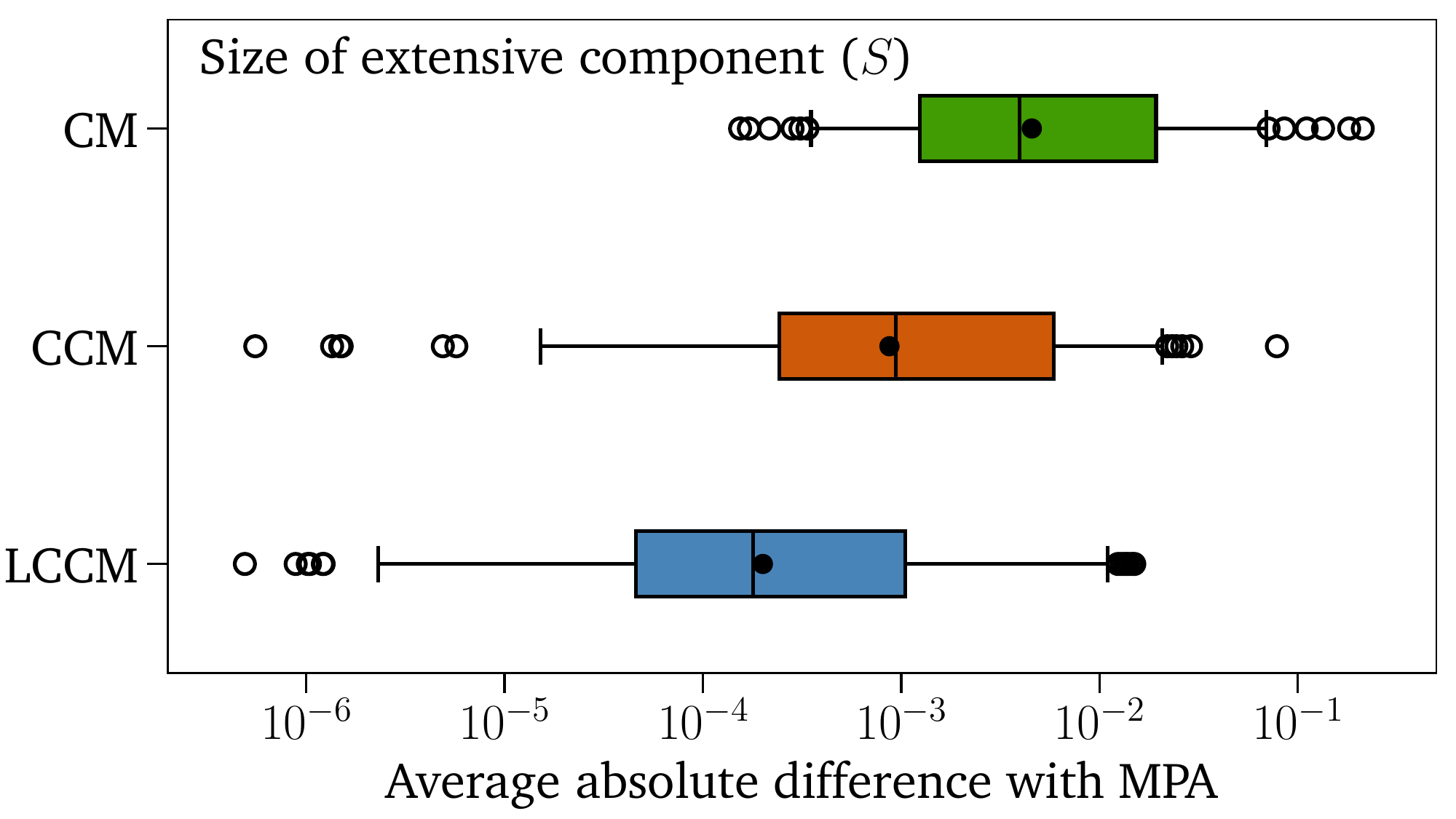}
  \caption{Predictions of the intensive models (CM, CCM and LCCM) compared to the predictions of the extensive MPA for 111 real biological, technological, transportation and social complex networks downloaded from \texttt{icon.colorado.edu}.  The whiskers cover the range between the 5th and the 95th percentiles, the black dots indicate the mean and the outliers data points are shown with a circle.  Each box indicates the first, second and third quartiles, as usual.
  (left) Relative error of the percolation threshold defined as $|p_\mathrm{c}^\mathrm{model}-p_\mathrm{c}^\mathrm{MPA}|/p_\mathrm{c}^\mathrm{MPA}$.  The calculation of $p_\mathrm{c}^\mathrm{LCCM}$ is detailed in Methods.
  (right) Area of the region bounded by the curves $S^\mathrm{model}$ and $S^\mathrm{MPA}$ computed as $\int_0^1 |S^\mathrm{model}-S^\mathrm{MPA}|dp$.
  References.~\cite{Newman2002,Vazquez2003,Karrer2014} provide the methods to compute $p_\mathrm{c}^\mathrm{model}$ and $S^\mathrm{model}$ for the CM, the CCM and the MPA.}
  \label{fig:thresholds}
\end{figure*}

To compute the size of the extensive component, we assume that the networks in the ensemble are locally tree-like, which occurs in the limit of large network size or when the detailed structure of matrix $\mathbf{L}$ only permits exact trees (i.e., when loops are structurally impossible).  We define $a_{lk}^\alpha$ as the probability that attempting to reach a node in class $(l,k)$ by one of its stubs of color $\alpha$ does not eventually lead to the extensive component.  Noting $p$ the probability that links are occupied, the probabilities $\{a_{lk}^\alpha\}$ are the solution of
\begin{align} \label{eq:a_lk_self_consistency}
  a_{lk}^\alpha = 1 - p + p f_{lk}^\alpha(\bm{a}) \ ,
\end{align}
for all $l$, $k$ and $\alpha$.  This last expression encodes the simple self-consistent argument that attempting to reach the node will not lead to the extensive component if 1) the link is unoccupied, which occurs with probability $1-p$, or if 2) the link is occupied, with probability $p$, but the attempts to reach the other neighbors of the node that has just been reached will all fail, which occurs with probability $f_{lk}^\alpha(\bm{a})$.  Note that this argument relies on the assumption that the state of these neighbors are independent, which is true for a tree-like structure.  Having solved Eq.~\eqref{eq:a_lk_self_consistency}, the relative size of the extensive component, $S$, is then given by the probability that a randomly chosen node is found in $S$
\begin{align}
  S = 1 - \sum_{lk} P(l,k) g_{lk}(\bm{a}) \ ,
\end{align}
where $P(l,k)$ is the fraction of nodes in class $(l,k)$ which can be extracted from the link correlation matrix $\mathbf{L}$ (see Methods).  Notice that since we assume the networks of the ensemble to be tree-like, the relative size of the extensive component if nodes (instead of links) were occupied with probability $p$ is simply $S^\mathrm{site} = pS$ to account for the probability that the initial randomly chosen node is occupied.  Note also that the percolation threshold, $p_\mathrm{c}$, is the value of $p$ at which $\bm{a}=\bm{1}$ becomes an unstable solution of Eq.~\eqref{eq:a_lk_self_consistency} (see Methods), which corresponds to the emergence of the extensive component.
%
%
%
%
%
\subsection*{Effective tree-like structure}
%
Because it is a subset of both the CM and the CCM, the cardinality of the ensemble defined by the LCCM should, in principle, be smaller than the ensembles considered by the formers.  Consequently, if the mesoscale structural information provided by the layers $l$ is of any significance, we expect the predictions of the LCCM to be the closest to the ones obtain with the MPA.  Figures~\ref{fig:bifurcation}~and~\ref{fig:thresholds} confirm this observation.  In fact, our results demonstrate that identifying nodes using the layer in the OD alongside their degree does not merely improve the predictions, it drastically changes their nature, making them qualitatively very similar to the ones of the MPA when not strikingly quantitatively identical.  As shown on Fig.~\ref{fig:bifurcation}, the LCCM reproduces the general shape of the curves, has the same number of inflection points, and always predict a connected network when all links are occupied (i.e., $S$ must be 1 at $p=1$ since we considered the largest connected components of every datasets). Interestingly, only the LCCM and the MPA are able to capture the mesoscopic core-periphery and/or modular structures that were numerically shown to lead to smeared (or double) phase transitions~\cite{Colomer-de-Simon2014} such as the one observed on the protein-protein interaction network.

Perhaps most importantly, the LCCM approximates to high accuracy the percolation threshold predicted by the MPA, as seen in Fig.~\ref{fig:thresholds}(left), with an relative error of less than 1.5\% for 75\% of the 111 network datasets considered.  Additionally, Fig.~\ref{fig:thresholds}(right) shows the expected error on the size of the extensive component averaged over the entire range of occupation probability $p$. When using the LCCM to compress the network structure, we find that the error, relative to the MPA, to be of the order of $10^{-3}$ for 75\% of the datasets considered; an improvement of at least one order of magnitude from existing approaches.  Altogether, these results indicate that categorizing nodes with the classes $(l,k)$ captures critical features of the local and mesoscopic tree-like organization of many real complex networks, thus offering an intensive effective description of their structure.
%
%
%
%
%
\section*{Conclusion}
%
We introduced a random network ensemble that relies solely on an \textit{intensive} description of the network structure that, nevertheless, yields predictions for percolation that are either essentially quantitatively identical---or at least strikingly qualitatively similar---to the ones obtained with the state-of-the-art MPA.  This ensemble assigns two structural features to each node---its degree $k$ (local) and its position $l$ in the Onion Decomposition of the network (mesoscale)---and creates links according to simple connection rules that exactly preserve these two features.  This ensemble lends itself to exact analytical calculations using probability generating functions in the limit of large network size, and is mathematically principled, meaning that it leads to exact predictions on trees, like the MPA, but unlike other intensive approaches such as the configuration model and its variants.  The accuracy of the predictions of the LCCM shows that the OD easily captures important features of the mesoscale structural organization of many real complex networks, and that this information should be leveraged by the future generations of models of complex networks.

For instance, Eq.~\eqref{eq:varphi_definition}, which provides the distribution of different link types (e.g., the number of links leading to lower or higher layers) for any node, could be straightforwardly included in equations for other problems such as the Susceptible-Infectious-Susceptible dynamics.  It would thus be possible to track the fraction of infected nodes with a given pair $(l,k)$ whose time evolution would be driven by the transmission events along the connections prescribed by Eqs.~\eqref{eq:g}--\eqref{eq:f}.  In a purely numerical context, and using a simpler, less accurate version of the LCCM, this approach was already shown to lead to predictions of SIS dynamics that are an order of magnitude more precise than other network models \cite{Hebert-Dufresne2016a}.  More generally, the pair $(l,k)$ consists in a straightforward and computationally inexpensive observable to characterize and rank nodes based on their local connectivity (through $k$) and global centrality (through $l$).

Finally, the accuracy of the LCCM strongly suggests that the long-range correlations induced by the OD effectively emulate the correlations considered in the MPA, and, consequently, that a large chunk of the structural properties behind the accuracy of the MPA now lend themselves to intensive analytical treatment.  This opens the way for future work to focus on bringing the analytical modeling of complex networks beyond the ubiquitous tree-like approximation. Doing so should provide a unified framework for random graphs, regular structures like lattices, and the complex networks that lie in-between.
%
%
%
%
%
%
%
%
%
%
%
%
\clearpage
\onecolumngrid
\section*{Methods}
%
%
%
%
%
%
\subsection*{Link correlation matrix}
%
We define the symmetrical link correlation matrix $\mathbf{L}$ whose elements, $L_{lk, l^\prime k^\prime}$, correspond to the fraction of links between nodes of class $(l,k)$ and $(l^\prime,k^\prime)$.  It has the following properties
\begin{align}
  \frac{1}{2} \sum_{lk} \sum_{l^\prime k^\prime} (1 + \delta_{ll^\prime}\delta_{kk^\prime}) L_{lk, l^\prime k^\prime} = 1 \ ,
\end{align}
since each type of links appears twice in the matrix except for the links connecting nodes of the same class (i.e., diagonal elements), and
\begin{align} \label{eq:links_vs_stubs}
  \frac{1}{2} \sum_{l^\prime k^\prime} (1 + \delta_{ll^\prime}\delta_{kk^\prime}) L_{lk, l^\prime k^\prime} = \frac{k P(l,k)}{\langle k \rangle} \ ,
\end{align}
where $P(l,k)$ is the fraction of nodes belonging to the class $(l,k)$ and $\langle k \rangle = \sum_{lk}kP(l,k)$ is the average degree.
%
%
%
%
%
\subsection*{Distribution of the number and of the color of stubs}
%
The connection rules of the LCCM indicate that a node of degree $k$ in layer $l$ and coreness $c_l$ have at most $c_l$ red stubs.  Since red stubs are defined as half-links toward nodes in layers $l^\prime \geq l$, they represent a fraction
\begin{align}
  \frac{1}{2}\sum_{l^\prime \geq l} \sum_{k^\prime} (1 + \delta_{ll^\prime}\delta_{kk^\prime}) L_{lk, l^\prime k^\prime}
\end{align}
of all stubs in the network ensemble, where $\delta_{ll^\prime}\delta_{kk^\prime}$ accounts for the fact that a link connecting two nodes of class $(l,k)$ contribute to two red stubs.  This last quantity would be equal to
\begin{align}
  \frac{c_l P(l,k)}{\langle k \rangle} 
\end{align}
if every of these nodes had exactly $c_l$ red stubs.  Consequently, since the LCCM only dictates bounds on the number of each color, the probability that a node of degree $k$ in layer $l$ has exactly $k^\mathrm{r}$ red stubs is simply
\begin{align} \label{eq:prob_number_red_stubs}
  \binom{c_l}{k^\mathrm{r}}& \big[ p_{lk}^{\mathrm{r}} \big]^{k^\mathrm{r}} \big[ 1 - p_{lk}^{\mathrm{r}} \big]^{c_l-k^\mathrm{r}}
\end{align}
where
\begin{align} \label{eq:prob_red}
   p_{lk}^\mathrm{r} & = \frac{\sum_{l^\prime \geq l} \sum_{k^\prime} (1 + \delta_{ll^\prime}\delta_{kk^\prime}) L_{lk, l^\prime k^\prime}}{2 c_l P(l,k)  / \langle k \rangle} \ .
\end{align}
Note that whenever layer $l$ is the first layer of its core---when $c_{l}>c_{l-1}$---Eq.~\eqref{eq:prob_red} reduces to $p_{lk}^{\mathrm{r}}=1$ meaning that each node has exactly $c_l$ red stubs, as prescribed by the connection rules of the LCCM.

Similarly, the fraction of half-links shared with nodes in layers $l^\prime<l-1$ (i.e., green stubs) is
\begin{align} \label{eq:fraction_green_stubs}
  \frac{1}{2} \sum_{l^\prime < l-1}  \sum_{k^\prime} L_{lk, l^\prime k^\prime} \ .
\end{align}
The maximal value of this quantity, however, varies in function of $l$.  If the layer is the first layer of its shell (i.e., if $c_{l}>c_{l-1}$), then each node has $c_l$ red stubs and up to $k-c_{l}$ green stubs according to the connections rules.  If $c_l=c_{l-1}$, nodes that have exactly $c_l$ red stubs can have up to $k-c_l-1$ green stubs since they must have at least one black stubs, and can have up to $k-c_l$ otherwise.  The maximal value of Eq.~\eqref{eq:fraction_green_stubs} can therefore be summarized as
\begin{align}
  \frac{(k - c_l - \delta_{k^\mathrm{r},c_l}\delta_{c_l,c_{l-1}}) P(l,k)}{\langle k \rangle} \ ,
\end{align}
such that the probability that a node of degree $k$ in layer $l$ has exactly $k^\mathrm{g}$ green stubs is
\begin{align} \label{eq:prob_number_green_stubs}
  \binom{k-c_l-\delta_{k^\mathrm{r},c_l}\delta_{c_l,c_{l-1}}}{k^\mathrm{g}} \left[ \frac{k-c_l}{k-c_l-\delta_{k^\mathrm{r},c_l}\delta_{c_l,c_{l-1}}} p_{lk}^{\mathrm{g}} \right]^{k^\mathrm{g}} \left[ 1 - \frac{k-c_l}{k-c_l-\delta_{k^\mathrm{r},c_l}\delta_{c_l,c_{l-1}}} p_{lk}^{\mathrm{g}} \right]^{k-c_l-k^\mathrm{g}-\delta_{k^\mathrm{r},c_l}\delta_{c_l,c_{l-1}}}
\end{align}
with
\begin{align}
  p_{lk}^\mathrm{g} = \frac{\sum_{l^\prime < l-1}  \sum_{k^\prime} L_{lk, l^\prime k^\prime}}{2(k - c_l) P(l,k)/\langle k \rangle} \ .
\end{align}
Combining Eqs.~\eqref{eq:prob_number_red_stubs} and \eqref{eq:prob_number_green_stubs} yields the probability that a node in layer $l$ and of degree $k$ has $k^\mathrm{r}$, $k^\mathrm{g}$ and $k^\mathrm{b}$ red, green and blacks stubs, respectively
\begin{align} \label{eq:stub_number_and_color_distribution}
  &P_{lk}(k^\mathrm{r},k^\mathrm{g},k^\mathrm{b}) = \displaystyle \delta_{k,k^\mathrm{r}+k^\mathrm{b}+k^\mathrm{g}} \binom{c_l}{k^\mathrm{r}} \big[ p_{lk}^{\mathrm{r}} \big]^{k^\mathrm{r}} \big[ 1 - p_{lk}^{\mathrm{r}} \big]^{c_l-k^\mathrm{r}} \nonumber \\
    & \times \binom{k-c_l-\delta_{k^\mathrm{r},c_l}\delta_{c_l,c_{l-1}}}{k^\mathrm{g}} \left[ \frac{k-c_l}{k-c_l-\delta_{k^\mathrm{r},c_l}\delta_{c_l,c_{l-1}}} p_{lk}^{\mathrm{g}} \right]^{k^\mathrm{g}} \left[ 1 - \frac{k-c_l}{k-c_l-\delta_{k^\mathrm{r},c_l}\delta_{c_l,c_{l-1}}} p_{lk}^{\mathrm{g}} \right]^{k-c_l-k^\mathrm{g}-\delta_{k^\mathrm{r},c_l}\delta_{c_l,c_{l-1}}} \ .
\end{align}
Finally, after some elementary algebra, it can be shown that the generating function $\varphi_{lk}(\bm{x})$ associated with this distribution is
\begin{align} \label{eq:varphi}
  \varphi_{lk}(\bm{x}) & = \sum_{k^\mathrm{r},k^\mathrm{b},k^\mathrm{g}} P(k^\mathrm{r},k^\mathrm{g},k^\mathrm{b}|l,k)
                          [x_{lk}^{\mathrm{r}}]^{k^\mathrm{r}} [x_{lk}^{\mathrm{g}}]^{k^\mathrm{g}} [x_{lk}^{\mathrm{b}}]^{k^\mathrm{b}} \nonumber \\
                       & = \delta_{c_l, c_{l-1}}
                           x_{lk}^\mathrm{b} \left[ p_{lk}^{\mathrm{r}} x_{lk}^{\mathrm{r}} \right]^{c_l}
                           \left[ \left(1 - \frac{k-c_l}{k-c_l-1} p_{lk}^{\mathrm{g}}\right) x_{lk}^{\mathrm{b}} + \frac{k-c_l}{k-c_l-1} p_{lk}^{\mathrm{g}} x_{lk}^{\mathrm{g}} \right]^{k-c_l-1} \nonumber \\
                       & \qquad\qquad\qquad\qquad\qquad
                         - \delta_{c_l,c_{l-1}}
                           \left[ p_{lk}^{\mathrm{r}} x_{lk}^{\mathrm{r}} \right]^{c_l}
                           \left[ \left(1 - p_{lk}^{\mathrm{g}}\right) x_{lk}^{\mathrm{b}} + p_{lk}^{\mathrm{g}} x_{lk}^{\mathrm{g}} \right]^{k-c_l} \nonumber \\
                       & \qquad\qquad\qquad\qquad\qquad
                         + \left[ (1 - p_{lk}^{\mathrm{r}}) x_{lk}^{\mathrm{b}} + p_{lk}^{\mathrm{r}} x_{lk}^{\mathrm{r}} \right]^{c_l}
                           \left[ (1 - p_{lk}^{\mathrm{g}}) x_{lk}^{\mathrm{b}} + p_{lk}^{\mathrm{g}} x_{lk}^{\mathrm{g}} \right]^{k-c_l} \ .
\end{align}
%
%
%
%
%
\subsection*{Transition probabilities}
%
With the distribution of the number of stubs of each color that nodes have being provided by Eq.~\eqref{eq:varphi}, the only missing quantities are the transition probabilities: the probability $Q_{lk}^\alpha(l^\prime,k^\prime,\alpha^\prime)$ that a stub of color $\alpha$ stemming from a node of class $(l,k)$ leads to a stub of color $\alpha^\prime$ attached to a node of class $(l^\prime,k^\prime)$.  Once more, this information can be extracted from the link correlation matrix $\mathbf{L}$.

Let us recall that black stubs stemming from nodes of class $(l,k)$ can only lead to red stubs attached to nodes in the previous layer (i.e., $l^\prime=l-1$), which can be summarized by
\begin{align} \label{eq:black_to_red_stubs}
  Q_{lk}^\mathrm{b}(l^\prime,k^\prime,\alpha^\prime) = \frac{\delta_{\alpha^\prime,\mathrm{r}} \delta_{l^\prime,l-1} L_{l^{\prime}k^{\prime},lk}}{\sum_{l^{\prime\prime}}\sum_{k^{\prime\prime}}\delta_{l^{\prime\prime},l-1}L_{l^{\prime\prime}k^{\prime\prime},lk}} \ ,
\end{align}
where the denominator is proportional to the fraction of all stubs that are black and that are stemming from nodes of class $(l,k)$.  Similarly, since green stubs can only lead to red stubs attached nodes in layer $l^\prime<l-1$, we have
\begin{align} \label{eq:green_to_red_stubs}
  Q_{lk}^\mathrm{g}(l^\prime,k^\prime,\alpha^\prime) =
  \left\{
  \begin{array}{cc}
    \displaystyle\frac{\delta_{\alpha^\prime,\mathrm{r}} L_{l^\prime k^\prime, lk}}{\sum_{l^{\prime\prime}<l-1} \sum_{k^{\prime\prime}} L_{l^{\prime\prime}k^{\prime\prime},lk}}  & \text{if } l^\prime < l-1 \\
    &\\
    0 & \text{otherwise}
  \end{array}\right. \ .
\end{align}
Because red stubs can lead to all three colors of stubs, we first consider the case where a red stubs leads to a black stubs (i.e., to a node in layer $l^\prime=l+1$), which corresponds to
\begin{align}
  Q_{lk}^\mathrm{r}(l^\prime,k^\prime,\mathrm{b}) = \frac{\delta_{l^\prime,l+1}L_{lk,l^\prime k^\prime}}{\sum_{l^{\prime\prime} \geq l} \sum_{k^{\prime\prime}} (1 + \delta_{ll^{\prime\prime}}\delta_{kk^{\prime\prime}}) L_{lk,l^{\prime\prime}k^{\prime\prime}}} \ ,
\end{align}
where the denominator is proportional to the fraction of all stubs that corresponds to red stubs stemming from nodes of class $(l,k)$.  In the case of red stubs leading to red stubs---i.e., links between nodes in the same layer---, we need to double the contribution of $L_{lk,lk}$ since each link between nodes of the same class contributes to two red stubs, which yields

\begin{align}
  Q_{lk}^\mathrm{r}(l^\prime,k^\prime,\mathrm{r}) = \frac{\delta_{ll^\prime}(1 + \delta_{kk^{\prime\prime}}) L_{lk,l^{\prime\prime}k^{\prime\prime}}}{\sum_{l^{\prime\prime} \geq l} \sum_{k^{\prime\prime}} (1 + \delta_{ll^{\prime\prime}}\delta_{kk^{\prime\prime}}) L_{lk,l^{\prime\prime}k^{\prime\prime}}} \ .
\end{align}
The case of red stubs leading to green stubs is similar to Eq.~\eqref{eq:green_to_red_stubs} and is straightforward to obtain
\begin{align} \label{eq:red_to_green_stubs}
  Q_{lk}^\mathrm{r}(l^\prime,k^\prime,\mathrm{g}) =
  \left\{
  \begin{array}{cc}
    \displaystyle\frac{L_{lk, l^\prime k^\prime}}{\sum_{l^{\prime\prime} \geq l} \sum_{k^{\prime\prime}} (1 + \delta_{ll^{\prime\prime}}\delta_{kk^{\prime\prime}}) L_{lk,l^{\prime\prime}k^{\prime\prime}}}   & \text{if } l^\prime > l+1 \\
    &\\
    0 & \text{otherwise}
  \end{array}
  \right. \ .
\end{align}
Finally, by injecting Eqs.~\eqref{eq:black_to_red_stubs}--\eqref{eq:red_to_green_stubs} in Eq.~\eqref{eq:gamma}, we obtain

\begin{subequations}
\begin{align}
  \gamma_{lk}^\mathrm{r}(\bm{x}) & = \frac{\sum_{l^{\prime} \geq l} \sum_{k^{\prime}} L_{lk,l^{\prime}k^{\prime}}
  [\delta_{ll^\prime} (1 + \delta_{kk^{\prime}}) x_{l^\prime k^\prime}^\mathrm{r} + \delta_{ll^\prime-1}x_{l^\prime k^\prime}^\mathrm{b} + (1 - \delta_{ll^\prime})(1 - \delta_{ll^\prime-1})x_{l^\prime k^\prime}^\mathrm{g}]}{\sum_{l^{\prime\prime} \geq l} \sum_{k^{\prime\prime}} (1 + \delta_{ll^{\prime\prime}}\delta_{kk^{\prime\prime}}) L_{lk,l^{\prime\prime}k^{\prime\prime}}} \label{eq:gammas_r} \\
  \gamma_{lk}^\mathrm{b}(\bm{x}) & = \frac{\sum_{k^\prime}L_{l-1k^\prime,lk}x_{l-1k^\prime}^\mathrm{r}}{\sum_{k^{\prime\prime}}L_{l-1k^{\prime\prime},lk}} \\
  \gamma_{lk}^\mathrm{g}(\bm{x}) & = \frac{\sum_{l^\prime<l-1} \sum_{k^\prime} L_{l^\prime k^\prime,lk}x_{l^\prime\mathrm{r}}^\mathrm{r}}{\sum_{l^{\prime\prime}<l-1} \sum_{k^{\prime\prime}} L_{l^{\prime\prime}k^{\prime\prime},lk}} \label{eq:gammas_g} \ .
\end{align}
\end{subequations}
%
%
%
%
%
\subsection*{Percolation threshold}
%
The value of the percolation threshold, $p_\mathrm{c}$, can be computed analytically by a linear stability analysis of the solution $\bm{a}=\bm{1}$ of Eq.~\eqref{eq:a_lk_self_consistency}.  Substituing $a_{lk}^\alpha = 1 - \varepsilon_{lk}^\alpha$, where $\varepsilon_{lk}^\alpha \ll 1$, yields
\begin{align}
  \varepsilon_{lk}^{\alpha} = p \sum_{l^\prime k^\prime \alpha^\prime} \left. \frac{\partial f_{lk}^{\alpha}(\bm{x})}{\partial x_{l^\prime k^\prime}^{\alpha^\prime}} \right|_{\bm{x}=\bm{1}} \varepsilon_{l^\prime k^\prime}^{\alpha^\prime} \ ,
\end{align}
when limiting the expansion of $ f_{lk}^{\alpha}(\bm{1} - \bm{\varepsilon})$ to the first order.  The last equation can be rewritten as an eigenvalue problem
\begin{align}
  \bm{\varepsilon} = p \mathbf{M} \bm{\varepsilon} \ ,
\end{align}
thus indicating that the fixed point $\bm{a}=\bm{1}$ looses its stability---i.e., the extensive component emerges---when the largest eigenvalue of $p\mathbf{M}$ exceeds 1.  The percolation threshold, $p_\mathrm{c}$, therefore equals the reciprocal of the largest eigenvalue of $\mathbf{M}$ which, by virtue of the Perron-Frobenius theorem, is real and positive.

The elements of $\mathbf{M}$ can be written as
\begin{align}
  \frac{\partial f_{lk}^{\alpha}(\bm{1})}{\partial x_{l^\prime k^\prime}^{\alpha^\prime}}
    = \frac{1}{\langle k^\alpha \rangle_{lk}} \sum_{\alpha^{\prime\prime}} \frac{\partial^2 \varphi_{lk}(\bm{1})}{\partial x_{lk}^{\alpha} \partial x_{lk}^{\alpha^{\prime\prime}}} \frac{\partial \gamma_{lk}^{\alpha^{\prime\prime}}(\bm{1})}{\partial x_{l^\prime k^\prime}^{\alpha^\prime}} \ ,
\end{align}
where the derivatives are calculated directly from Eq.~\eqref{eq:varphi} and Eqs.~\eqref{eq:gammas_r}--\eqref{eq:gammas_g}.  While the derivatives of $\gamma_{lk}^\alpha\bm{x}$ are straightforward, the derivatives of $\varphi_{lk}(\bm{x})$ require special care with respect to the value of $k-c_l$.  To facilitate the numerical implementation of the formalism, we provide the explicit expression of the derivatives of $\varphi_{lk}(\bm{x})$.
\begin{subequations}
\begin{align}
  \langle k^\mathrm{r} \rangle_{lk} & = \frac{\partial \varphi_{lk}(\bm{1})}{\partial x_{lk}^\mathrm{r}} = c_l p_{lk}^\mathrm{r} \\
  \langle k^\mathrm{g} \rangle_{lk} & = \frac{\partial \varphi_{lk}(\bm{1})}{\partial x_{lk}^\mathrm{g}} = 
  \left\{
  \begin{array}{lc}
    (k - c_l) p_{lk}^\mathrm{g} - \delta_{c_l, c_{l-1}} [p_{lk}^\mathrm{r}]^{c_l} (k - c_l) p_{lk}^\mathrm{g} & \text{if } k - c_l \leq 1 \\
    &\\
    (k - c_l) p_{lk}^\mathrm{g}                                                                  & \text{otherwise}
  \end{array}\right. \\
  \langle k^\mathrm{b} \rangle_{lk} & = \frac{\partial \varphi_{lk}(\bm{1})}{\partial x_{lk}^\mathrm{b}} =
  \left\{
  \begin{array}{lc}
    c_l (1 - p_{lk}^\mathrm{g}) + (k - c_l) (1 - p_{lk}^\mathrm{g}) \\
    \qquad\qquad +\ \delta_{c_l, c_{l-1}} [p_{lk}^\mathrm{r}]^{c_l} (k - c_l) p_{lk}^\mathrm{g} & \text{if } k - c_l \leq 1 \\
    &\\
    c_l (1 - p_{lk}^\mathrm{r}) + (k - c_l) (1 - p_{lk}^\mathrm{g})                                                                                      & \text{otherwise}
  \end{array}\right.
\end{align}
\begin{align}
  \frac{\partial^2 \varphi_{lk}(\bm{1})}{\partial x_{lk}^{\mathrm{r}\,2}} & = c_l (c_l - 1) [p_{lk}^\mathrm{r}]^2 \\
\nonumber \\
  \frac{\partial^2 \varphi_{lk}(\bm{1})}{\partial x_{lk}^{\mathrm{r}}\partial x_{lk}^{\mathrm{g}}} & =
  \left\{
  \begin{array}{lc}
  c_l (k - c_l) p_{lk}^\mathrm{r} p_{lk}^\mathrm{g} \\
  \qquad\qquad -\ \delta_{c_l,c_{l-1}} [p_{lk}^\mathrm{r}]^{c_l} c_l (k - c_l) p_{lk}^\mathrm{g} & \text{if } k - c_l \leq 1 \\
  & \\
  c_l (k - c_l) p_{lk}^\mathrm{r} p_{lk}^\mathrm{g}                                              & \text{otherwise}
  \end{array} \right. \\
\nonumber \\
  \frac{\partial^2 \varphi_{lk}(\bm{1})}{\partial x_{lk}^{\mathrm{r}}\partial x_{lk}^{\mathrm{b}}} & =
  \left\{
  \begin{array}{lc}
  c_l (c_l - 1) p_{lk}^\mathrm{r} (1 - p_{lk}^\mathrm{r}) + c_l (k - c_l) p_{lk}^\mathrm{r} (1 - p_{lk}^\mathrm{g}) \\
  \qquad\qquad +\ \delta_{c_l,c_{l-1}} [p_{lk}^\mathrm{r}]^{c_l} c_l (k - c_l) p_{lk}^\mathrm{g} & \text{if } k - c_l \leq 1 \\
  & \\
  c_l (c_l - 1) p_{lk}^\mathrm{r} (1 - p_{lk}^\mathrm{r}) + c_l (k - c_l) p_{lk}^\mathrm{r} (1 - p_{lk}^\mathrm{g})                                              & \text{otherwise}
  \end{array} \right.\\
\nonumber \\
  \frac{\partial^2 \varphi_{lk}(\bm{1})}{\partial x_{lk}^{\mathrm{g}\,2}} & =
  \left\{
  \begin{array}{lc}
  (k - c_l) (k - c_l - 1) [p_{lk}^\mathrm{g}]^2 \\
  \qquad\qquad -\ \delta_{c_l, c_{l-1}} [p_{lk}^\mathrm{r}]^{c_l} (k - c_l) (k - c_l - 1) [p_{lk}^\mathrm{g}]^2 & \text{if } k - c_l \leq 2 \\
  & \\
  (k - c_l) (k - c_l - 1) [p_{lk}^\mathrm{g}]^2 \\
  \qquad\qquad -\ \delta_{c_l, c_{l-1}} [p_{lk}^\mathrm{r}]^{c_l} (k - c_l) (k - c_l - 1) [p_{lk}^\mathrm{g}]^2 \\
  \qquad\qquad +\ \delta_{c_l, c_{l-1}} [p_{lk}^\mathrm{r}]^{c_l} (k - c_l)^2 [p_{lk}^\mathrm{g}]^2 \displaystyle \frac{k - c_l - 2}{k - c_l -1}& \text{otherwise}
  \end{array} \right. \\
\nonumber \\
  \frac{\partial^2 \varphi_{lk}(\bm{1})}{\partial x_{lk}^{\mathrm{g}}\partial x_{lk}^{\mathrm{b}}} & =
  \left\{
  \begin{array}{lc}
  c_l (k-c_l) p_{lk}^\mathrm{g} (1 - p_{lk}^\mathrm{r})   & \text{if } k - c_l \leq 1 \\
  & \\
  c_l (k-c_l) p_{lk}^\mathrm{g} (1 - p_{lk}^\mathrm{r}) + (k - c_l) (k - c_l - 1) p_{lk}^\mathrm{g} (1 - p_{lk}^\mathrm{g}) \\
  \qquad\qquad +\ \delta_{c_l, c_{l-1}} [p_{lk}^\mathrm{r}]^{c_l} (k - c_l) (k - c_l - 1) [p_{lk}^\mathrm{g}]^2 & \text{if } k - c_l = 2 \\
  & \\
  c_l (k-c_l) p_{lk}^\mathrm{g} (1 - p_{lk}^\mathrm{r}) + (k - c_l) (k - c_l - 1) p_{lk}^\mathrm{g} (1 - p_{lk}^\mathrm{g}) \\
  \qquad\qquad +\ \delta_{c_l, c_{l-1}} [p_{lk}^\mathrm{r}]^{c_l} (k - c_l) (k - c_l - 1) [p_{lk}^\mathrm{g}]^2 \\
  \qquad\qquad -\ \delta_{c_l, c_{l-1}} [p_{lk}^\mathrm{r}]^{c_l} (k - c_l)^2 [p_{lk}^\mathrm{g}]^2 \displaystyle\frac{k - c_l - 2}{k - c_l - 1} & \text{otherwise}
  \end{array} \right. \\
\nonumber \\
  \frac{\partial^2 \varphi_{lk}(\bm{1})}{\partial x_{lk}^{\mathrm{b}\,2}} & =
  \left\{
  \begin{array}{lc}
  c_l (c_l - 1) (1 - p_{lk}^\mathrm{r})^2 + 2 c_l (k - c_l) (1 - p_{lk}^\mathrm{r}) (1 - p_{lk}^\mathrm{g}) & \text{if } k - c_l \leq 1 \\
  & \\
  c_l (c_l - 1) (1 - p_{lk}^\mathrm{r})^2 + 2 c_l (k - c_l) (1 - p_{lk}^\mathrm{r}) (1 - p_{lk}^\mathrm{g}) \\
  \qquad\qquad +\ (k - c_l) (k - c_l - 1) (1 - p_{lk}^\mathrm{g})^2 \\
  \qquad\qquad -\ \delta_{c_l, c_{l-1}} [p_{lk}^\mathrm{r}]^{c_l} (k - c_l) (k - c_l - 1) [p_{lk}^\mathrm{g}]^2 & \text{if } k - c_l = 2 \\
  & \\
  c_l (c_l - 1) (1 - p_{lk}^\mathrm{r})^2 + 2 c_l (k - c_l) (1 - p_{lk}^\mathrm{r}) (1 - p_{lk}^\mathrm{g}) \\
  \qquad\qquad +\ (k - c_l) (k - c_l - 1) (1 - p_{lk}^\mathrm{g})^2 \\
  \qquad\qquad -\ \delta_{c_l, c_{l-1}} [p_{lk}^\mathrm{r}]^{c_l} (k - c_l) (k - c_l - 1) [p_{lk}^\mathrm{g}]^2 \\
  \qquad\qquad +\ \delta_{c_l, c_{l-1}} [p_{lk}^\mathrm{r}]^{c_l} (k - c_l)^2 [p_{lk}^\mathrm{g}]^2 \displaystyle\frac{k - c_l - 2}{k - c_l - 1} & \text{otherwise}
  \end{array} \right.
\end{align}
\end{subequations}
Let us recall that $c_l \neq c_{l-1}$ and $p_{lk}^\mathrm{r} = 1$ whenever $k = c_l$ since these nodes are in the first layer of their core by definition, and that we set $c_1 \neq c_0$ to simplify the notation.  Note also that
\begin{align}
  \sum_{\alpha} \frac{\partial \varphi_{lk}(\bm{1})}{\partial x_{lk}^\alpha} =
  \sum_{\alpha} \langle k^\alpha \rangle_{lk} = k
  \qquad \text{ and } \qquad
  \sum_{\alpha, \alpha^\prime} \frac{\partial^2 \varphi_{lk}(\bm{1})}{\partial x_{lk}^\alpha \partial x_{lk}^{\alpha^\prime}} =
  \sum_{\alpha, \alpha^\prime} \langle k^\alpha (k^{\alpha^\prime} - \delta_{\alpha\alpha^\prime}) \rangle_{lk} = k(k - 1)
\end{align}
for $\alpha, \alpha^\prime \in \{\mathrm{r}, \mathrm{g}, \mathrm{b}\}$ and regardless of the value of $k-c_l$, as expected.
\end{document}